\documentclass[pre,showpacs,amsfonts,amssymb,eqsecnum,%
twocolumn,floatfix,longbibliography,superscriptaddress]{revtex4-1}
\usepackage{amsmath}
\usepackage{graphicx}
\usepackage{bm}
\allowdisplaybreaks

\begin{document}
	
\title{Demographic stochasticity and extinction in populations with Allee effect}
	
	\author{Vicen\c{c} M\'{e}ndez}
	\affiliation{Grup de F{\'i}sica Estad\'{i}stica.  Departament de F{\'i}sica.
		Facultat de Ci{\`e}ncies. Edifici Cc. Universitat Aut\`{o}noma de Barcelona,
		08193 Bellaterra (Barcelona) Spain}
	\author{Michael Assaf}
	\affiliation{Racah Institute of Physics, Hebrew University of Jerusalem,
		Jerusalem 91904, Israel}
	\author{Axel Mas\'{o}-Puigdellosas}
		\affiliation{Grup de F{\'i}sica Estad\'{i}stica.  Departament de F{\'i}sica.
		Facultat de Ci{\`e}ncies. Edifici Cc. Universitat Aut\`{o}noma de Barcelona,
		08193 Bellaterra (Barcelona) Spain}
	\author{Daniel Campos}
		\affiliation{Grup de F{\'i}sica Estad\'{i}stica.  Departament de F{\'i}sica.
		Facultat de Ci{\`e}ncies. Edifici Cc. Universitat Aut\`{o}noma de Barcelona,
		08193 Bellaterra (Barcelona) Spain}
	\author{Werner Horsthemke}
	\affiliation{Department of Chemistry, Southern Methodist University,
		Dallas, Texas 75275-0314, USA}
	\date{\today}
	
	\begin{abstract}
We study simple stochastic scenarios, based on birth-and-death Markovian processes, that describe populations with Allee effect, to account for the role of demographic stochasticity. In the mean-field deterministic limit we recover well-known deterministic evolution equations widely employed in population ecology. The mean-time to extinction is in general obtained by
the Wentzel-Kramers-Brillouin (WKB) approximation for populations with strong and weak Allee effects. An exact solution for the mean time to extinction can be found via a recursive equation for special cases of the stochastic dynamics. We study the conditions for the validity of the WKB solution and analyze the boundary between the weak and strong Allee effect by comparing exact solutions with numerical simulations.
	\end{abstract}
	
	\maketitle

\section{Introduction}

The Allee effect describes situations in population ecology where the per capita population growth rate displays a hump-like shape.
The growth rate decreases both at low densities as the population density decreases, and at high densities as the population increases
\cite{Cou08,CoClGr99,Ste99}.
Various mechanisms can cause a declining per capita growth rate for declining population density at low densities \cite{De89,CoClGr99,Ste99b,Be06}.
The Allee effect can be due to factors that affect reproduction, such as the difficulty of finding a mate at low population densities or the fact that small breeding groups are less successful in producing and rearing young. It can also be caused by factors that affect survival, such as less efficient defense against predators by small populations or the ability to locate food. The Allee effect is called strong, if it results in a negative per capita growth rate once the population size falls below a threshold, leading to deterministic extinction. The Allee effect is called weak if there is no threshold and the per capita growth rate is small but remains positive even at very low population densities.
It has been suggested that demographic stochasticity represents a mechanism for the Allee effect \cite{La98}, although other authors argue that the fluctuations in a population with a low number of individuals do not lead to a reduced individual fitness and hence cannot be considered as a mechanism of the Allee effect.

In this paper we study an isolated population that undergoes a set of stochastic Markovian  birth-and-death processes based on individual interactions or on effective gain-loss processes that account for a shortage of interactions among members of small populations.
In this way, we provide  two
stochastic models, both of which demonstrate the strong as well as the weak Allee effect. In the deterministic mean-field limit our models give rise to rate equations for the population dynamics that are widely employed in the literature \cite{De89,CoClGr99,Ste99b,De16}.

We focus on determining the mean time to extinction (MTE), the mean time it takes the system to become extinct due to demographic fluctuations, when starting from the vicinity of the deterministic stable state. In the deterministic limit, when demographic fluctuations are ignored, once the system reaches the deterministic stable state it stays there forever. However, since the system is isolated, demographic fluctuations ultimately drive the system to extinction, see, e.g., Refs.~\cite{nisbet2003modelling,lande2003stochastic,allen2010introduction}.

The standard approach for finding the MTE is based on the Fokker-Planck or Kolmogorov approximation to the master equation, where the population size is assumed to be a continuous variable. While it is still being used to deal with environmental and demographic fluctuation in population extinction \cite{De12}, it has been shown that the Fokker-Planck approximation fails to correctly estimate the MTE for large fluctuations that take the population size far away from the deterministic stable state \cite{Gaveau1996,DoSaSa05,Ts14,assaf2006spectral,assaf2007spectral,AsMe10,assaf2010large}.

Recently, an alternative approach to find the MTE, based on the
Wentzel-Kramers-Brillouin (WKB) approximation for the
master equation, has been developed, which provides accurate results for general birth-and-death processes, see, e.g., Refs.~\cite{Dyk94,kessler2007extinction,meerson2008noise,EsKa09,AsMe10,assaf2017wkb}. We employ the WKB approach to find the MTE for both the weak and strong Allee effect.  In the special case of \textit{single-step} birth-and-death processes, see below, one can solve a recursive equation, derived directly from the master equation,  to obtain an exact solution for the MTE.  Studying these special birth-and-death processes has the
significant advantage that we can compare the exact solution with the result from the WKB approximation and determine the range of validity of the latter. To the best of our knowledge, the MTE in the case of the weak Allee effect has not been determined before. We explore how the MTE depends on characteristic parameters of the population and study specifically the boundary region between the weak and strong Allee effect.

The paper is organized as follows. In Sec.\ \ref{sec:detAllee} we discuss two widely used deterministic models for populations with Allee effect. We introduce an individual-based birth-and-death model in Sec.\ \ref{sec:IBM} and determine the MTE for the weak and strong Allee effect via the WKB approximation. In Sec.\ \ref{sec:mshort} we study a phenomenological model with birth-and-death processes whose transition rates are density-dependent and obtained on phenomenological grounds or from empirical data. We again use the WKB approximation to obtain the MTE for the weak and strong Allee effect.  We present the exact solution for the MTE via a recursive relation for the special case of single-step birth-and-death processes  in Sec.\ \ref{sec:exact}. This result is used in Sec.\ \ref{sec:valid} to establish the range of validity, and the quality, of the WKB approximation. Sec.\ \ref{sec:thresh} is devoted to studying the boundary between the weak and strong Allee effect, where the linear birth and linear death rates are equal. We discuss our results
in Sec.\ \ref{sec:concl}.

\section{Deterministic models for the Allee effect}\label{sec:detAllee}

Many works dealing with single-species populations that
display the Allee effects are based on simple phenomenological
or empirical, non-spatial, deterministic models. The essence
of these models goes back to the original work by Odum and Allee \cite{OdAl54},
where the observed per capita growth rate was fitted by a suitable
function. A general form for a deterministic model in continuous time
is the following differential equation,
\begin{equation}
\frac{d\rho(t)}{dt}=f(\rho),\label{eq:ode}
\end{equation}
where $\rho(t)$ is the average population size (average number of individuals) at time
$t$, and $f(\rho)$ is a function specifying the form of the effective growth rate of the population at size $\rho$. Many types of functions $f(\rho)$ have been proposed
in the literature. A review of different deterministic
models can be found in Ref.\ \cite{BoBe02}.

\subsection{Cubic model}
The simplest
and oldest model goes back to Volterra's paper \cite{Vo38}, where $f(\rho)$ is a cubic
polynomial function of $\rho$. This model is based on the fact that for a constant ratio of males and females the number of meetings between the two sexes is proportional to $\rho^{2}$. The model takes into account that the ratio of births and meetings can be affected by the population density and  assumes it to be a linearly decreasing function of $\rho$. In addition, the model includes birth and mortality events that occur at constant per capita rates.
The Volterra model has the following form,
\begin{equation}
\frac{d\rho(t)}{dt}=-a_1\rho+(a_2-a_3\rho)\rho^{2}=-a_1\rho+a_2\rho^{2}-a_3\rho^{3},
\label{eq:volterra}
\end{equation}
where $a_2$ and $a_3$ are positive parameters. The parameter
 $a_1$ can be either positive or negative, depending on the difference
 between the linear birth and mortality rates. If one defines the two
real-valued roots,
\begin{subequations}
\label{k12}
\begin{align}
k_{1}&=\frac{1}{2a_3}\left[a_2-\sqrt{a_2^{2}-4a_1a_3}\right],\\
k_{2}&=\frac{1}{2a_3}\left[a_2+\sqrt{a_2^{2}-4a_1a_3}\right],
\end{align}
\end{subequations}
with $a_2^2>4a_1a_3$, then the model is often cast in the form \cite{CoClGr99}
\begin{equation}
\frac{d\rho(t)}{dt}=a_1\rho\left(1-\frac{\rho}{k_{2}}\right)\left(\frac{\rho}{k_{1}}-1\right),
\label{eq:v2}
\end{equation}
which emphasizes its similarity to the logistic
equation with a new unstable steady state, $\rho=k_{1}$.

\begin{figure}[htbp]
	\includegraphics[width=0.97\hsize]{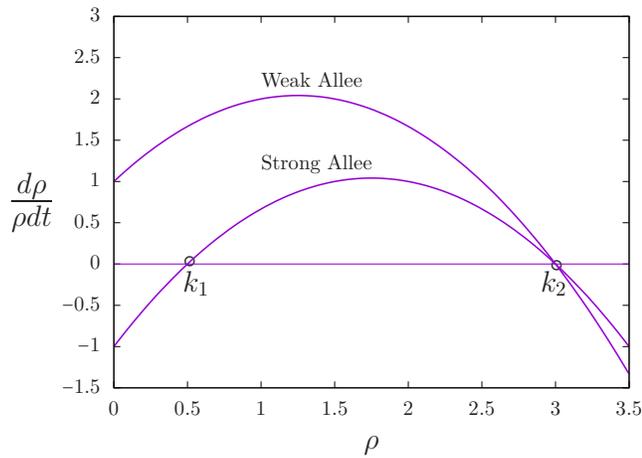}
	\caption{A schematic plot of the per capita growth rate for the Volterra model. The growth rate is always positive in the case of the weak Allee effect. It is negative below the critical density $k_1$ in the case of the strong Allee effect.}
	\label{fig:f1}
\end{figure}

Based on the value of $a_1$ there exist two dynamical scenarios. If
$a_1>0$, the Volterra model has three steady states, two stable states at $\rho=0$ and $\rho=k_2$ and an unstable state at $\rho=k_1$. Here, if the initial population size is larger than
$k_{1}$, the population increases in time and converges
to the stable steady state $\rho=k_{2}$, the carrying capacity of the system.
If the initial population size is smaller than $k_{1}$, the population decays to the stable steady state $\rho=0$ and becomes
extinct. This scenario describes the so-called strong Allee effect.
A different dynamical scenario occurs when $a_1<0$. Here $k_{1}$ becomes negative and the steady state $\rho=k_{1}$
lacks biological meaning. In this case the Volterra model has
only two steady states, an unstable state at $\rho=0$ and a
stable state at $\rho=k_{2}$. The population growth rate is positive
but smaller than for the logistic equation. This case corresponds
to the so-called weak Allee effect.
In Fig.~\ref{fig:f1} we plot the per capita growth rate $\rho^{-1}d\rho/dt$
as a function of $\rho$ to allow a clear visualization and distinction between these two Allee scenarios.

\subsection{Logistic model with mating shortage}
\begin{figure}[htbp]
	\includegraphics[width=\hsize]{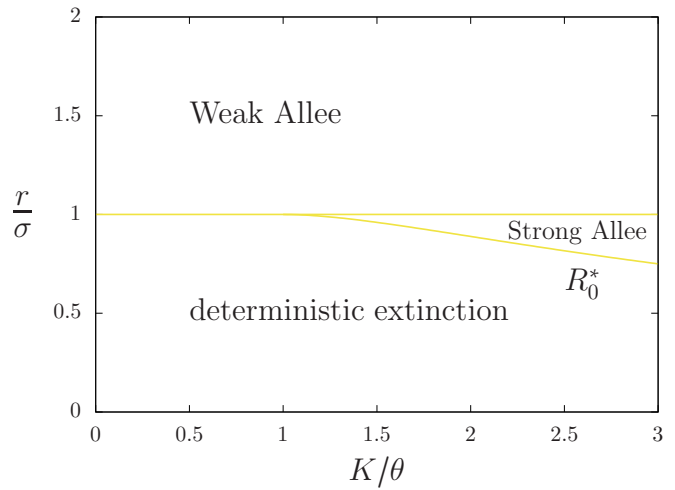}
	\caption{Parameter space of the logistic model with mating shortage described by the rate equation~(\ref{romod2}).}
	\label{fig:f2}
\end{figure}

An alternative way to account for the Allee effect is to
add a term involving the frequency of mating encounters in the population   to the logistic equation. Several functional forms have been considered \cite{De89}, and we chose
the rectangular hyperbola model for the encounter function, $\mathcal{P}$, corresponding to the probability of mating
\begin{equation}
\mathcal{P}=\frac{\rho}{\rho+\theta}\label{matp}.
\end{equation}
Here $\theta$ is the population size at which the probability
of mating is 1/2.
The greater the value of $\theta$, the harder it is to find a mate at low population densities, giving rise to an Allee effect in the population.
The probability of not mating is \cite{De89}
\begin{equation}
\bar{\mathcal{P}}=1-\mathcal{P}=\frac{\theta}{\rho+\theta}.
\label{eq:notm}
\end{equation}
A term proportional to $\bar{\mathcal{P}}$ is subtracted from the
logistic equation to account for the reduction in offspring due to
mating shortage,
and thus, the logistic growth with Allee effect can be described by
\cite{De89,CoClGr99,Ste99b}
\begin{equation}
\frac{d\rho(t)}{dt}=r\rho\left(1-\frac{\rho}{K}\right)-\frac{\sigma\theta\rho}{\rho+\theta}.\label{romod2}
\end{equation}
Here $r$ is the intrinsic growth rate, $K$ the carrying capacity,
and $\sigma$ measures the magnitude or severity of the Allee effect.
Equation  (\ref{romod2}) has a steady
state, $\rho=0$, i.e., extinction, which is stable for $r<\sigma$ and unstable for $r>\sigma$, and two other steady states
$\rho_{\pm}$,
\begin{equation}
\rho_{\pm}=\frac{K-\theta}{2}\left[1\pm\sqrt{1-\frac{4K\theta(\sigma-r)}{r(K-\theta)}}\,\right]>0,\nonumber
\end{equation}
provided that $K>\theta$. The stability of these states and the sign of
the growth rate $d\rho/dt$ for small $\rho$ are similar to those of the Volterra equation, again giving rise to the scenarios of the strong and weak Allee effect. In Fig.\ \ref{fig:f2} we summarize the different possible
situations depending on the values of the parameters. The region corresponding to the strong Allee effect lies between $r/\sigma=1$ and $r/\sigma=R_0^{*}$, with
\begin{equation}
R_0^{*}=\frac{4K}{(1+K/\theta)^2}.
\label{R0*}
\end{equation}

\section{Stochastic individual-based model with Allee effect}\label{sec:IBM}

To study the effects of internal fluctuations or demographic stochasticity,
we follow the usual practice and assume that the temporal evolution of the population is given by a Markovian birth-and-death process. Two approaches are commonly employed in the literature to obtain the transition rates for the Markovian process. In individual-based models (IBM) of populations, one lists the specific birth, death, and competition processes that individuals in the population experience. These processes naturally determine the transition rates, in the same way as the elementary reactions of a chemical reaction system do. In the second approach, the transition rates are based on phenomenological considerations or empirical data, without any explicit reference to the underlying processes that the individuals experience. We consider an IBM in this section and a phenomenological model in Sec.\ \ref{sec:mshort}.

The minimal individual-based model that displays both the weak and strong Allee effect consists of two birth processes (linear and binary birth), a ternary competition process, and a linear death process. It can be expressed as a chemical reaction system:
\begin{subequations}
	\label{gsys1}
	\begin{align}
		\text{X} & \xrightarrow{\mu} (1+b)\,\text{X},\\
		2\,\text{X} & \xrightarrow{\lambda}  (2+a)\,\text{X},\\
\text{X} & \xrightarrow{\gamma} \emptyset,\\
	3\,\text{X} & \xrightarrow{\beta}  (3-c)\,\text{X}.
	\end{align}
\end{subequations}
The first reaction is a linear birth process, which occurs at a constant rate $\mu$, and describes the base-line reproductive success of the population without any cooperative effects. It accounts for the fact that individuals produce $b$ offspring which reach reproductive age. The second reaction is a binary process and occurs at a constant rate $\lambda$. It describes cooperative interactions, such as cooperative breeding, cooperative anti-predator behavior, or cooperative foraging, that result in producing $a$ additional offspring which reach reproductive age.  The third reaction is a linear death process, at constant rate $\gamma$, which accounts for mortality due to natural causes. The last reaction is a ternary competition process, describing the effects of overcrowding, resource depletion, etc., where $c$ individuals die at rate $\beta$. Only the values $c=1,2,3$ are meaningful.

Two particular cases of system~\eqref{gsys1}, which do \textit{not} display an Allee effect, were recently analyzed. When $\lambda=0$, the second reaction vanishes and the theta-logistic equation is obtained in the mean field limit~\cite{MeAsHo17}. When $\beta=0$, the last reaction vanishes and the logistic equation is recovered in the mean field limit \cite{MeAsCa15}.

The reaction scheme \eqref{gsys1}
defines a Markovian birth-and-death process, and the temporal evolution of $P(n,t)$, the probability of having $n$ individuals at time $t$, is described by the following master equation, also known as the forward Kolmogorov equation~\cite{Ga90}:
\begin{equation}
\frac{d P(n,t)}{d t}=\sum_{r}
\left[W(n-r,r)P(n-r,t)-W(n,r)P(n,t)\right],
\label{eq:me}
\end{equation}
where it is understood that $P(n<0,t)=0$. Here $W(n,r)$ are
the transition rates between the states with $n$ and $n+r$ individuals, where $r=\{r_{1},r_{2},r_{3},r_{4}\}=\{a,b,-1,-c\}$
are the transition increments corresponding to the system given by Eqs.\ (\ref{gsys1}). The transition rates
corresponding to each reaction, $W(n,r)$, are obtained from the reaction kinetics \cite{Ga90,vK07,HoLe84}:
\begin{subequations}
\label{trans}
\begin{align}
	W(n,a)&=\frac{\lambda}{2}n(n-1),\\
	W(n,b)&=\mu n,\\
	W(n,-1)&=\gamma n,\\
	W(n,-c)&=\frac{\beta}{6}n(n-1)(n-2).
\end{align}
\end{subequations}

Deterministic mean-field equations for the expected or average population size can be obtained directly from (\ref{eq:me}). Multiplying Eq.~(\ref{eq:me}) by $n$, using transition rates (\ref{trans}), and summing over all values of $n$, we find
\begin{equation}
\frac{d\rho}{dt}=(\mu b-\gamma)\rho +\frac{a\lambda }{2}\rho^{2}-\frac{c\beta}{6}\rho^{3},\label{eq:mfe1}
\end{equation}
where $\rho = \langle n\rangle$ is the mean number of individuals.
This deterministic equation strictly holds when the demographic fluctuations vanish, which occurs when the population size goes to infinity. Equation~(\ref{eq:mfe1}) can be cast in the form of Eq.~(\ref{eq:v2}) with the definitions
\begin{subequations}
\label{k122}
\begin{align}
k_{1}&=\frac{3}{2c\beta}\left[a\lambda-\sqrt{a^{2}\lambda^2+8c\beta(b\mu-\gamma)/3}\,\right],\\
k_{2}&=\frac{3}{2c\beta}\left[a\lambda+\sqrt{a^{2}\lambda^2+8c\beta(b\mu-\gamma)/3}\,\right],
\end{align}
\end{subequations}
where we have used Eq.~(\ref{k12}). At the deterministic level, the interaction scheme~\eqref{gsys1} of the IBM gives rise to the Volterra rate equation, Eq.~(\ref{eq:v2}), which is still used to study Allee effects \cite{CoClGr99}.

For the sake of simplicity we focus in the following
on the simplest version of this IBM, namely $a=b=c=1$.
The set of interactions (\ref{gsys1}) becomes
\begin{subequations}
	\label{gsys}
	\begin{align}
		\text{X} & \xrightarrow{\mu} 2\,\text{X},\\
		2\,\text{X} & \xrightarrow{\lambda}  3\,\text{X},\\
		\text{X} & \xrightarrow{\gamma} \emptyset,\\
		3\,\text{X} & \xrightarrow{\beta}  2\,\text{X}.
	\end{align}
\end{subequations}
The mean-field rate equation corresponding to (\ref{gsys}) is
\begin{equation}
\frac{d\rho}{dt}=(\mu -\gamma)\rho +\frac{\lambda }{2}\rho^{2}-\frac{\beta}{6}\rho^{3}.
\label{mf2}
\end{equation}
For this set of reactions, the master equation can be obtained by substituting (\ref{trans}) with $a=b=c=1$ into (\ref{eq:me}), which yields
\begin{multline}
\frac{d P(n,t)}{d t}=(n-1)\left[\frac{\lambda}{2}(n-2)+\mu  \right]P(n-1,t)\\
+(n+1)\left[\frac{\beta}{6}n(n-1)+\gamma \right]P(n+1,t)\\
-n\left[\frac{\lambda}{2}(n-1)+\mu +\frac{\beta}{6}(n-1)(n-2)+\gamma \right]P(n,t).
\label{eq:me30}
\end{multline}

The master equation (\ref{eq:me30}) includes only single-step processes where the transitions take place between the states $n$ and $n\pm 1$.
Defining the new dimensionless quantities in terms of the reaction rates
\begin{equation}
N=\frac{3\lambda}{2\beta},\quad\delta^2=1+\frac{8\beta (\mu-\gamma)}{3\lambda^2},\quad R_0=\frac{\mu}{\gamma},
\label{rel}
\end{equation}
we can write the steady states of Eq.\ (\ref{mf2}) simply as
\begin{subequations}
\label{eqst}
\begin{align}
n_e&=0,\\
n_e^u&=N(1-\delta),\\
n_e^s&=N(1+\delta).
\end{align}
\end{subequations}
Here the superscripts $u$ and $s$ stand for unstable and stable, respectively, while the stability of the state $n_e=0$ depends on whether the Allee effect is weak or strong, see below. Note that $N$ defines the scale of the typical population size prior to extinction.
The identities (\ref{rel}) establish a relation between the microscopic ($\lambda$, $\mu$, $\gamma$, and $\beta$) and
macroscopic ($N$, $\delta$, $R_0$) parameters, which can be obtained from field observations.
If we compare Eqs.\ (\ref{mf2}) and (\ref{eq:volterra}), we realize that the IBM displays  both types of Allee effects:
\begin{subequations}
\label{dis}
\begin{align}
\text{Weak Allee:} \quad &\mu>\gamma\;\;\text{or}\;\;R_0>1,\\
\text{Strong Allee:} \quad &\mu<\gamma\;\;\text{or}\;\;R_0<1.
\end{align}
\end{subequations}
Note that for $R_0>1$ ($R_0<1$) we have $\delta>1$ ($\delta < 1$), where for the strong Allee effect, we must also demand that $\delta>0$, see below. Also note that, while in the regime of the strong Allee effect, the steady state $n_e=0$ is stable, for the weak Allee effect, $\delta>1$, the fixed point $N(1-\delta)$ lacks biological meaning, and the steady state $n_e=0$ is unstable.

We now consider both scenarios of weak and strong Allee effect and calculate the MTE when $\delta \neq 1$ and $R_0\neq 1$. To determine the MTE we employ the WKB approach. While the general procedure is the same for both scenarios, some steps of the derivation of the MTE are different for each case. We summarize here the main aspects of the derivation. Further details can be found in Refs.\ \cite{EsKa09,AsMe10,assaf2017wkb}. The WKB approach provides an accurate result for the MTE when $N$ is sufficiently large. We consider the case where the population is well established and long lived, and is not in immediate danger of extinction, which requires that the typical population size, $n_e^s$ [see Eq.~(\ref{eqst})], be large. We assume henceforth that $N\gg 1$, while $\delta$ is assumed to be $O(1)$. This scaling is fulfilled by choosing a scaling of the parameters, such that $\mu=O(1)$, $\gamma=O(1)$,
$\lambda =O(N^{-1})$, and $\beta =O(N^{-2})$, which is naturally imposed by the overall orders of the various reaction steps.

We begin by rescaling time $t\to \gamma t$ and introducing the rescaled population number density $q=n/N$, where $N$ is given in (\ref{rel}).
We now follow the general formalism outlined in
Refs.~\cite{EsKa09,AsMe10,assaf2017wkb}, and expand the transition rates in powers of $N\gg 1$, as
\begin{equation}
W(Nq,r_i)=Nw_{r_i}(q)+u_{r_i}(q)+O(N^{-1}),
\label{eq:exp}
\end{equation}
where $w_{r_i}(q)$ and $u_{r_i}(q)$ are ${O}(1)$ for $q={O}(1)$, and $i=1,2,3,4$. Making use of Eq.~(\ref{trans}) with $a=b=c=1$ and Eq.~(\ref{rel}), we have
\begin{subequations}
\label{eq:tr}
\begin{align}
w_{r_1=1}&=\frac{2(R_0-1)}{\delta^2-1}q^2,\\
 u_{r_1=1}&=-\frac{2(R_0-1)}{\delta^2-1}q,\\
w_{r_2=1}&=R_{0}q,\\
 u_{r_2=1}&=0,\\
w_{r_3=-1}&=q,\\
 u_{r_3=-1}&=0,\\
w_{r_4=-1}&=\frac{R_0-1}{\delta^2-1}q^3,\\
 u_{r_4=-1}&=-3\frac{R_0-1}{\delta^2-1}q^2.
\end{align}
\end{subequations}
Using these rates, we can define the leading-order total single-step birth and death rates,
\begin{subequations}
\label{wpm}
\begin{align}
w_{+}(q)&=w_{r_1=1}+w_{r_2=1}=\frac{2(R_0-1)}{\delta^2-1}q^2+R_0q,\\
w_{-}(q)&=w_{r_3=-1}+w_{r_4=-1}=\frac{R_0-1}{\delta^2-1}q^3+q,
\end{align}
\end{subequations}
while the subleading-order corrections become
\begin{subequations}
\label{upm}
\begin{align}
u_{+}(q)&=u_{r_1=1}+u_{r_2=1}=-\frac{2(R_0-1)}{\delta^2-1}q,\\
u_{-}(q)&=w_{r_3=-1}+w_{r_4=-1}=-3\frac{R_0-1}{\delta^2-1}q^2.
\end{align}
\end{subequations}

A necessary condition for extinction is that  $w_{\pm}(0)=u_{\pm}(0)=0$, indicating
that the extinction state is an absorbing state. This occurs in this model regardless of the parameter values, and thus, demographic fluctuations always drive the population experiencing Allee effects to extinction with a finite MTE.

\subsection{MTE for the case of the weak Allee effect}

In the case of the weak Allee effect, $\delta>1$, and the deterministic model (\ref{eq:v2}) or (\ref{eq:mfe1}) has only two positive steady states: $n_e=0$ and $n_e^s=N(1+\delta)$. The former is unstable and the latter is stable. In the presence of fluctuations, however, the stable steady state is only metastable. Extinction inevitably occurs after a finite time due to the presence of an absorbing state at $n=0$. Nevertheless,  the MTE is expected to be exponentially large for
$N\gg 1$~\cite{assaf2006spectral,assaf2007spectral,kessler2007extinction,assaf2010large,AsMe10,assaf2017wkb}, see below. Here the path to extinction follows a special instanton-like trajectory connecting  the stable and unstable steady states~\cite{Dyk94,kessler2007extinction,AsMe10}. In order to  calculate the MTE, we assume
that  the system has already converged to the long-lived metastable state $n_e^s$. As a result, the time-dependent probability distribution of population sizes can be approximately written as $P(n,t)\simeq
 \pi(n)e^{-t/\tau}$, where $\pi(n)$ is the quasi-stationary distribution (QSD) that describes the shape of the metastable state, while $\tau$ is the MTE~\cite{Dyk94,kessler2007extinction,AsMe10,assaf2017wkb}. By employing the WKB approximation for the QSD,
 it has been shown in Refs.~\cite{AsMe10,assaf2017wkb}
 that for generic single-step processes the MTE is given by
\begin{equation}
\tau=\frac{\sqrt{2\pi R_0 w'_{-}(q_e^s)} }{\gamma (R_0-1)\sqrt{Nw'_{+}(q_e^s)Q_s}}\,e^{N\Delta S+\Delta \phi},
\label{metwkb}
\end{equation}
with
\begin{align}
\Delta S&=\int_{0}^{q_e^s}\ln \left[ \frac{w_{+}(q)}{w_{-}(q)}\right] dq, \label{deltaS1}\\
\Delta \phi &=  \int_{0}^{q_e^s}\left[\frac{u_{+}(q)}{w_{+}(q)}-\frac{u_{-}(q)}{w_{-}(q)} \right]dq,  \label{deltafi1}\\
Q_{s}&=w'_{-}(q_e^s)w_{+}(q_e^s)-w'_{+}(q_e^s)w_{-}(q_e^s). \label{Qs}
\end{align}
Here $q_e^s =n_e^s/N$, and the prime denotes differentiation with respect to $q$.  The result is valid as long as $N\gg 1$. With $q_e^s =1+\delta$ and the specific birth and death rates given by Eqs.~(\ref{wpm}) and (\ref{upm}), we finally obtain
\begin{equation}
\tau_W=\frac{1}{\gamma}\sqrt{\frac{\pi(\delta-1)}{N\delta (\delta +1)}}\left( \frac{R_0}{R_0-1}\right)^{3/2}\,e^{N\Delta S},
\label{metwkb2}
\end{equation}
where
\begin{multline}
\Delta S=\frac{R_0\left( \delta^2\ -1 \right) }{2(R_0-1)} \ln \left[ \frac{R_0(\delta +1)-2}{R_0(\delta -1)}\right]+1+\delta\\
-2\sqrt{\frac{\delta^2-1}{R_0-1}}\arctan \left[ \sqrt{\frac{(R_0-1)(1+\delta)}{\delta-1}}\right].
\label{ds}
\end{multline}

\subsection{MTE for the case of the strong Allee effect}
In the case of the strong Allee effect, $\delta<1$, and there are three steady states, namely $q_e^s=n_e^s/N=1+\delta$ and $q_e=0$, which are stable, and $q_e^u=n_e^u/N=1-\delta$, which is unstable. The path to extinction follows an instanton connecting $q_e^s$ to $q_e^u$, from which the system flows to the extinction state almost deterministically~\cite{Dyk94,EsKa09,AsMe10}. Again, by employing the WKB approximation for the QSD, it has been shown in this case that the MTE is given by~\cite{EsKa09,AsMe10,assaf2017wkb}
\begin{equation}
\tau=\frac{2\pi}{\gamma} \sqrt{\frac{w_{-}(q_e^u)w_{+}(q_e^u)}{Q_s |Q_u|}}\, e^{N\Delta S+\Delta \phi},
\label{metwkbs}
\end{equation}
where
\begin{align}
\Delta S&=\int_{q_e^u}^{q_e^s}\ln \left[ \frac{w_{+}(q)}{w_{-}(q)}\right] dq,\\
\Delta \phi &=   \int_{q_e^u}^{q_e^s}\left[\frac{u_{+}(q)}{w_{+}(q)}-\frac{u_{-}(q)}{w_{-}(q)} \right]dq, \\
Q_u&= w'_{-}(q_e^u)w_{+}(q_e^u)-w'_{+}(q_e^u)w_{-}(q_e^u) \label{Qu},
\end{align}
and $Q_s$ is given by Eq.~(\ref{Qs}).

With $q_e^u =1-\delta$, $q_e^s =1+\delta$, and the specific birth and death rates given by~(\ref{wpm}) and (\ref{upm}), Eq.~(\ref{metwkbs}) yields
\begin{equation}
\tau_S = \frac{\pi (1-\delta)}{\gamma \delta(1-R_0)}e^{N\Delta S},
\label{metwkbs1}
\end{equation}
where
\begin{multline}
\Delta S= 2\delta -2\sqrt{\frac{1-\delta^2}{1-R_0}}\arctan\left(\frac{2\delta}{2-R_0}\sqrt{\frac{1-R_0}{1-\delta^2}}\right) \\
+\frac{R_0\left(1-\delta^2\right)}{2(1-R_0)}\ln
\left\{
	\frac{(1+\delta)[2-R_0(1+\delta)]}{(1-\delta)[2-R_0(1-\delta)]    }
	\right\}.
\label{metwkbs2}
\end{multline}
This result is well defined as long as $\delta<1$ and $R_0<1$. In addition, since $\delta^2$ must be positive,  Eq.~(\ref{rel}) implies that the rates must satisfy
\begin{equation}
\gamma <\mu+\frac{3\lambda ^2}{8\beta}.
\label{cond}
\end{equation}
Note that if condition (\ref{cond}) is not fulfilled, rate equation~(\ref{eq:mfe1}) does not display any Allee effect and has only one steady state: the extinction state. In that case, extinction occurs deterministically. As a result, Eq.~(\ref{cond}) is the condition for fluctuation-driven extinction.

\section{Stochastic model with mating shortage}\label{sec:mshort}

Instead of an individual-based model we consider in this section
a phenomenological
Markovian birth-and-death process with density-dependent transition rates. The dependence of the rates on the number of individuals has been proposed in \cite{No82,MaKi99,Du00,Ke00b,Naa01,AlAl03,McNe05,BlMc11,ChMc13} on
phenomenological grounds or by fitting them to empirical data \cite{De89}.
The model can be described by the following birth and death processes,
\begin{subequations}
	\label{gsysmodel2}
	\begin{align}
	\text{X} & \xrightarrow{W_{+}(n) } \text{X}+1,\\
	\text{X} & \xrightarrow{W_{-}(n)} \text{X}-1,
	\end{align}
\end{subequations}
where
\begin{subequations}
\label{model2}
\begin{align}
W_{+}(n)&=W(n,+1)=rn,\\
W_{-}(n)&=W(n,-1)=\frac{rn^{2}}{K}+\frac{\sigma\theta n}{\theta+n}.
\end{align}
\end{subequations}
The birth rate corresponds to linear Malthusian growth, and the
death rate takes into account competition for resources, through the
carrying capacity $K$, and the Allee term $\sigma\theta n/(\theta+n)$, corresponding to the probability of not mating~\cite{De89}, see Eq.\ \eqref{eq:notm} and the discussion below it.

We have added the Allee term to the death rate, since it acts like an effective mortality in the deterministic equation \eqref{romod2}. An alternative would be to subtract it from the birth rate,
\begin{subequations}
\label{model2a}
\begin{align}
W_{+}(n)&=W(n,+1)=rn-\frac{\sigma\theta n}{\theta+n},\\
W_{-}(n)&=W(n,-1)=\frac{rn^{2}}{K}.
\end{align}
\end{subequations}
The transition rate $W_{+}(n)$ must be nonnegative for $n=0,1,2,\dotsc$, which requires
that
\begin{equation}
\label{eq:nonneg}
r-\frac{\sigma\theta}{\theta+n}\geq 0\quad\text{for $n=1,2,\dotsc$}.
\end{equation}
Since $\sigma\theta/(\theta+n)$ is largest at $n=1$, and since $\theta=O(N)\gg 1$, condition~(\ref{eq:nonneg}) is fulfilled if and only if
\begin{equation}
\frac{r}{\sigma}\geq\frac{\theta}{\theta+1}=1-\frac{1}{\theta}+O(1/\theta^{2}).
\end{equation}
This implies that model \eqref{model2a} displays a strong Allee effect, which occurs for $r/\sigma <1$, see Fig.\ \ref{fig:f2}, only in a vanishingly small region in parameter space, namely $1-1/\theta+O(1/\theta^{2})< r/\sigma <1$. It is for this reason that we use model \eqref{model2}.

The deterministic mean-field equation for both models satisfies
$d\rho/dt=W_+(\rho)-W_-(\rho)$ and coincides with Eq. (\ref{romod2}).
As a result, the nonzero deterministic stable and unstable states  are $n_e^s=N(1+\delta)$ and $n_e^u=N(1-\delta)$, respectively, where
\begin{equation}
\delta^2=1-\frac{4s}{(s-1)^2}\left(\frac{1}{R_0}-1\right)
\label{delta2}
\end{equation}
and
\begin{equation}\label{rel2}
R_{0}=\frac{r}{\sigma},\quad N=\frac{K-\theta}{2}>0,\quad s=\frac{K}{\theta}>1.
\end{equation}

\subsection{MTE for the weak and strong Allee effect}
We now repeat the calculations performed in Sec.~\ref{sec:IBM} for the phenomenological model, and calculate the MTE for both cases of weak and strong Allee effect.
Rescaling time by $t\to t/\sigma$, introducing the number density $q=n/N$, defining $w_{\pm}(q)=W_{\pm}(n)/N$, and using Eq.~(\ref{rel2}), the birth and death rates (\ref{model2})
become
\begin{subequations}
\label{model2rates}
\begin{align}
w_{+}(q)&=R_{0}q,\\
w_{-}(q)&=R_{0}\left(\frac{s-1}{2s}\right)q^{2}+\frac{q}{1+q\left(\frac{s-1}{2}\right)}.
\end{align}
\end{subequations}
Since our model (\ref{gsysmodel2}) again includes only single-step processes, the MTEs are given by Eqs.~(\ref{metwkb}) and (\ref{metwkbs}) in the cases of weak and strong Allee effect, respectively. In the case of weak Allee effect we make use of Eqs.~(\ref{metwkb}) -- (\ref{Qs}), where $q_{e}^{s}=1+\delta$ and $r$ and $\sigma$ take the place of $\mu$ and $\gamma$, respectively. Therefore, the MTE for $R_{0}>1$ is
\begin{equation}
\tau_W=\frac{2e^{N\Delta S}}{\sigma(R_0-1)(1+\delta)(K-\theta)}\sqrt{\frac{\pi K}{R_0\delta}\left(\frac{K+\theta}{K-\theta}+\delta\right)},
\label{tauwfen}
\end{equation}
where $\delta$ is given by (\ref{delta2}) and $\Delta S=\int_{0}^{1+\delta}\ln(w_{+}/w_{-})dq$.

In the case of strong Allee effect the MTE is given by Eqs.\ (\ref{metwkbs}) -- (\ref{Qu}), where $q_{e}^{u}=1-\delta$ and $q_{e}^{s}=1+\delta$. As a result, the MTE for $R_{0}^{*}<R_{0}<1$ is
\begin{equation}
\tau_S=\frac{2\pi rK}{\delta(1+\delta)(K-\theta)}\left(\frac{K+\theta}{K-\theta}+\delta\right)e^{N\Delta S},
\label{tausfen}
\end{equation}
where $\Delta S=\int_{1-\delta}^{1+\delta}\ln(w_{+}/w_{-})dq$. For $R_0<R_0^*$, extinction is deterministic, see Fig.~\ref{fig:f2}.
Note that $\Delta S$ in both Eqs.\ (\ref{tauwfen}) and (\ref{tausfen})
can be calculated analytically in terms of elementary functions. However, the expressions are highly cumbersome and are therefore not displayed.

\section{Exact solution for the MTE}\label{sec:exact}

No general exact analytical solutions are known for the MTE for general multi-step Markovian birth-and-death processes, and one has to
employ approximation methods such as the WKB approach~\cite{AsMe10,assaf2017wkb}. This is the case for the general IBM \eqref{gsys1}. However, the simplest version of that model, namely \eqref{gsys} studied here, and the phenomenological birth-and-death process \eqref{model2} correspond to single-step birth-and-death processes.  For this class of processes, the MTE can be found analytically by solving a recursive equation. This feature allows us
to assess the range of validity and the quality of the WKB approximation. It also allows us to study the effects of demographic stochasticity in populations with Allee effect where the WKB approach cannot be used, namely when the linear birth and death rates, $\mu$ and $\gamma$, are too close to each other as shown below.

The master equation for the probabilities $P(n,t)$ for both models, \eqref{gsys} and \eqref{model2}, can be written in
the form
\begin{align}
\frac{dP(n,t)}{dt}&=w_+(n-1)P(n-1,t)+w_-(n+1)P(n+1,t)\nonumber\\
&-\left[w_+(n)+w_-(n)\right]P(n,t),
\label{eq:me2}
\end{align}
where $w_{+}(n)$ and $w_{-}(n)$ are the density-dependent
transition rates for the birth and death processes.
Since this master equation corresponds
to a single-step process, the MTE, $\tau(m)$, starting at a
state with $m$ individuals obeys the following recursive equation~\cite{Ga90}
\begin{multline}
w_{+}(m)\tau(m+1)+w_{-}(m)\tau(m-1)\\
-\left[w_{+}(m)+w_{-}(m)\right]\tau(m)=-1.
\label{eq:rec}
\end{multline}
We solve this recursive equation for both cases of the weak and strong Allee effect, assuming that the initial number of individuals $m$
is in the close vicinity of the stable deterministic steady state.
Since Eq.\ (\ref{eq:rec}) is a second-order difference equation, it is supplemented by two boundary conditions: (i) an absorbing boundary at $m=0$, i.e., $\tau(0)=0$, and (ii) a reflecting boundary at $m=\infty$, i.e., $\tau'(\infty)=0$~\cite{Ga90}.

We briefly outline the solution of Eq.~(\ref{eq:rec}), see details in Ref.~\cite{Ga90}. With the definition
\begin{equation}\label{eq:alpha}
\alpha(k)=\tau(k)-\tau(k-1), \quad k=1,2,3,\dotsc,
\end{equation}
Eq.~(\ref{eq:rec}) turns into
a first-order difference equation for $\alpha(k)$,
\begin{equation}
\alpha(k+1)-\frac{w_{-}(k)}{w_{+}(k)}\alpha(k)=-\frac{1}{w_{+}(k)}.\label{eq:alpha1}
\end{equation}
Further, by defining $p(k+1)=w_{-}(k)/w_{+}(k)$ and dividing Eq.~(\ref{eq:alpha1}) by $\prod_{j=1}^{k+1}p(j)$,
we can write Eq.~(\ref{eq:alpha1}) as
\begin{equation}
A(k+1)=A(k)-\varphi(k),
\label{eq:A}
\end{equation}
where
\begin{equation}
A(k)=\frac{\alpha(k)}{\prod_{j=1}^{k}p(j)},\quad\varphi(k)=\frac{1}{w_{-}(k)}\prod_{j=1}^{k}\frac{w_{+}(j-1)}{w_{-}(j-1)}.\label{eq:def}
\end{equation}
The solution of Eq.\ (\ref{eq:A}) is straightforward and can be written as
\begin{equation}
A(k)=A_{1}-\sum_{i=1}^{k-1}\varphi(i),\label{eq:As}
\end{equation}
where $A_{1}$ is a constant that is determined from the boundary
conditions. Since $\tau'(\infty)=0$, we have $\alpha(\infty)=A(\infty)=0$, and thus $A_{1}=\sum_{i=1}^{\infty}\varphi(i)$. As a result, Eq.~(\ref{eq:As}) reads
\begin{equation}
A(k)=\sum_{i=k}^{\infty}\varphi(i).\label{eq:Ak1}
\end{equation}
Finally, since the solution of Eq.~(\ref{eq:alpha}) satisfies $\tau (m)=\sum_{k=1}^{m}\alpha (k)$, by using Eqs.~(\ref{eq:def}) and (\ref{eq:Ak1}), we find after some algebra
\begin{equation}
\tau(m)=\sum_{k=1}^{m}\left\{ \frac{1}{w_{-}(k)}+\sum_{i=k+1}^{\infty}\frac{1}{w_{-}(i)}\prod_{j=0}^{i-1}\frac{w_{+}(j+k-1)}{w_{-}(j+k-1)}\right\}\!.
\label{taum}
\end{equation}
We remind the reader that in order to compare this result with the WKB approximation, the initial number of individuals $m$ has to be in the close vicinity of the stable fixed point. For simplicity we chose $m = \lfloor N(1+\delta)\rfloor$, where $\lfloor x\rfloor =\text{greatest integer $\leq x$}$.

\begin{figure}[htbp]
	\includegraphics[width=\hsize]{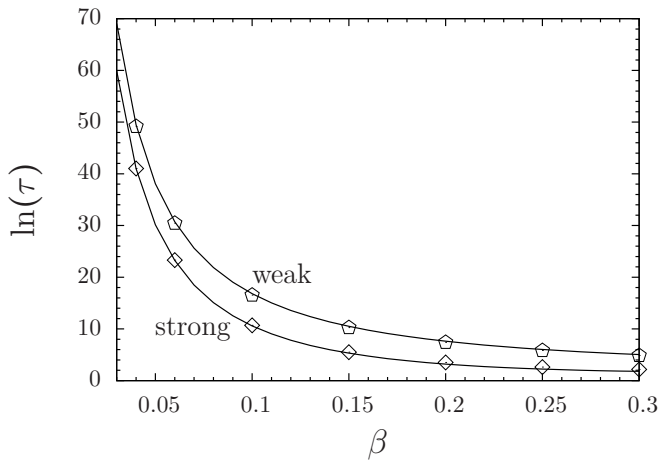}
	\caption{The logarithm of the MTE for the IBM is plotted as a function $\beta$, with $\lambda =\gamma=1$. The results of the weak Allee effect correspond to $\mu =1.5$, such that $R_0=1.5$, whereas the results of the strong Allee effect correspond to $\mu=0.3$, such that $R_0=0.3$. The solid lines are the WKB solutions given by Eqs.~(\ref{metwkb2}) and (\ref{ds}), and by Eqs.~(\ref{metwkbs1}) and (\ref{metwkbs2}), in the cases of the weak and strong Allee effect, respectively. Symbols denote the exact solution (\ref{taum}).}
	\label{fig:f3}
\end{figure}

In Fig.~\ref{fig:f3} we plot the MTE obtained for the IBM model for the cases of the weak and strong Allee effect, given by Eqs.~(\ref{metwkb2}) and (\ref{metwkbs1}), respectively, for different values of the competition
rate $\beta$. As expected, the MTE decreases as $\beta$ increases, since this causes an effective decrease in the typical system size $N$, see Eq.~(\ref{rel}). One can see that the WKB approximation for the MTEs in both cases of weak and strong Allee effect excellently agrees with the exact result~(\ref{taum}).

\begin{figure}[htbp]
	\includegraphics[width=\hsize]{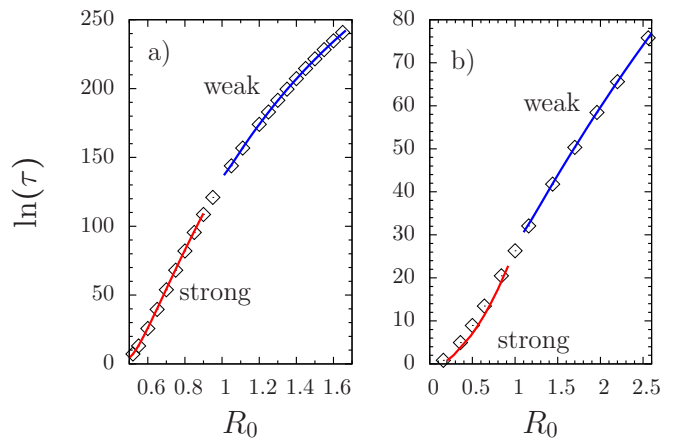}
	\caption{The logarithm of the MTE is plotted as a function of $R_0$. In panel (a) we plot the results for the phenomenological model for $K=600$, $\theta =100$ and $r =1$, such that $R_0^{*}=0.5$. Here we vary $\sigma$ in order to change $R_0$. The solid lines represent the WKB solutions given by (\ref{tauwfen}) and (\ref{tausfen}) above and below $R_0=1$, respectively, while the symbols denote the exact solution (\ref{taum}). In panel (b) we plot the results for the
IBM for $\gamma=25$, $\lambda=2$, $\beta=3/50$, and we vary $R_0$ by changing $\mu$. The solid lines correspond to the WKB solutions (\ref{metwkb2}) and (\ref{metwkbs1}) above and below $R_0=1$, respectively, while the symbols denote the exact solution (\ref{taum}). The WKB approximation breaks down when approaching $R_0=1$, which separates the regions of weak and strong Allee effect.}
	\label{fig:f4}
\end{figure}

In Fig.~\ref{fig:f4} we plot the MTE for both the IBM and the phenomenological model as function of the basic reproductive ratio $R_0$. As expected, the MTE increases with $R_0$. While for $R_0\neq 1$ (and not too close to $1$), the WKB solutions excellently agree with the exact results for the MTE in both models, as $R_0$ approaches $1$, the WKB solutions break down, see Fig.~\ref{fig:error} and below.

\section{Validity of the WKB solutions}\label{sec:valid}

The WKB approximation presented in the previous sections is valid as long as the typical population size is large. However, there is a further condition. We begin with the IBM \eqref{gsys} and show that in addition to $N\gg 1$, the linear birth and death rates, $\mu$ and $\gamma$, respectively, cannot be too close to each other. In other words, $R_0=\mu/\gamma$ [see Eq.~(\ref{rel})] cannot be too close to $1$.

In the case of the weak Allee effect, we need to find the
complete QSD, $\pi(n)$, to obtain
the WKB solution for the MTE. To do so, we need to match a recursive solution, valid at $1\ll n\ll \sqrt{N}$, and a WKB solution, valid at $n\gg 1$, for the QSD in their joint region of validity $1\ll n\ll \sqrt{N}$~\cite{AsMe10}. In order to find the lower bound on $R_0$ such that the WKB approximation is still valid, we introduce a small parameter $\varepsilon\!=\!R_0\!-\!1\!\ll\! 1$, and find its lower limit.

In Ref.~\cite{AsMe10} it was shown that the matching between the WKB and recursive solutions requires that $R_0^n\gg 1$ in the matching region. Evaluating $R_0^n$ at the upper limit of the matching region, $n=\sqrt{N}$,  we have $1\ll R_0^{\sqrt{N}}=e^{\sqrt{N}\ln R_0}\simeq e^{\sqrt{N}\varepsilon}$, since $\ln R_0=\ln(1+\varepsilon)\simeq \varepsilon$ for $\varepsilon\ll 1$. Therefore, in the case of the weak Allee effect, the WKB approximation is valid as long as $\varepsilon=R_0-1\gg N^{-1/2}$.

In the case of the strong Allee effect, finding the complete QSD requires matching the WKB solution, valid at $n-n_e^u\gg 1$, and a boundary-layer solution, valid at $\lvert n-n_e^u\rvert\ll N^{1/2}$, \textit{i.e.} in the close vicinity of the unstable fixed point $n_e^u$~\cite{EsKa09,AsMe10}. Since the matching is done for $1\ll \lvert n-n_e^u\rvert\ll N^{1/2}$, the distance between the unstable point and $n=0$ must be much larger than $N^{1/2}$, that is $n_e^u=N(1-\delta)\gg N^{1/2}$, or $1-\delta\gg N^{-1/2}$. Here $R_0<1$, and thus we define $\varepsilon=1-R_0\ll 1$. Since $\beta=O(N^{-2})$, see Sec.\ \ref{sec:IBM}, we write $\beta=B/N^2$, such that $B=O(1)$. Using Eq.~(\ref{rel}) we have $\lambda=2\beta N/3=2B/(3N)$, which yields $1-\delta\simeq 3\varepsilon\gamma/B$,  with $\gamma=O(1)$, see again Sec.\ \ref{sec:IBM}. This indicates that in the case of  the strong Allee effect, we must also have $\varepsilon\gg N^{-1/2}$ in order for the WKB approximation for the MTE to be accurate.

\begin{figure}[htbp]
	\includegraphics[width=\hsize]{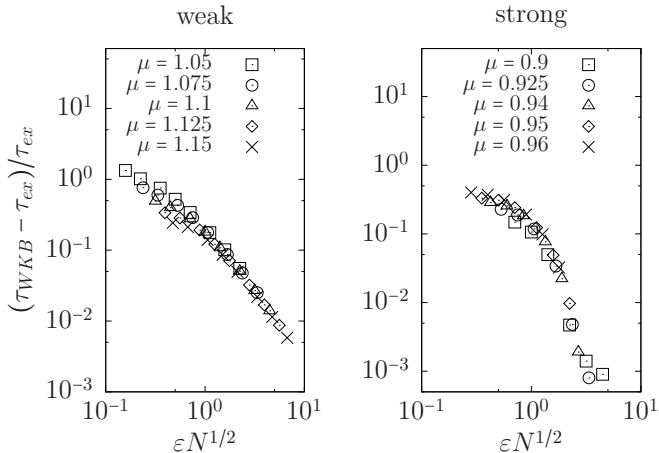}
	\caption{The relative error of the MTE, $(\tau_{WKB}-\tau_{ex})/\tau_{ex}$ is plotted for the IBM as a function of $\varepsilon N^{1/2}$ for the cases of weak (left panel) and strong (right panel) Allee effect, for five different values of $\mu$, see the legends. The WKB solution is given by Eqs.~(\ref{metwkb2}) and (\ref{ds}) in the case of weak Allee effect, and by Eqs.~(\ref{metwkbs1}) and (\ref{metwkbs2}) in the case of the strong Allee effect, while the exact solution is given by Eq.~(\ref{taum}). The parameters are $\gamma=1$, and $\lambda=2/(3N)$ with $N$ varying. Moreover, $\mu=1+\varepsilon$ and $\beta=1/N^2$ in the case of the weak Allee effect, while $\mu=1-\varepsilon$ and $\beta=3/N^2$, in the case of the strong Allee effect.}
	\label{fig:error}
\end{figure}

To verify this condition, we have plotted in Fig.~\ref{fig:error} the relative error of the MTE given by the ratio of $\tau_{WKB}-\tau_{ex}$ and $\tau_{ex}$, where $\tau_{ex}$ is the exact result. The left and right panels show the relative error of the MTE as function of $\varepsilon N^{1/2}$ for the cases of weak and strong Allee effects, respectively. According to our theoretical arguments presented above, we expect the relative error to be small once $\varepsilon N^{1/2}$ becomes large and vice versa, and this is exactly what is demonstrated in the figure.

In the phenomenological model, the linear birth rate is $r$ and the linear death rate for small populations is given by $\sigma$, such that one can define $\varepsilon=(r-\sigma)/\sigma$. In the weak Allee regime, the same criterion holds, namely $\varepsilon=R_0-1\gg 1/\sqrt{N}$. In the strong Allee regime, assuming that $s=O(1)$ [since both $K$ and $\theta$ scale as $O(N)$], and defining $\varepsilon=1-R_0$, we again obtain from the strong inequality $1-\delta\gg 1/\sqrt{N}$ that $\varepsilon\gg 1/\sqrt{N}$. Therefore, the general criterion for both models can be expressed as
\begin{equation}
\lvert 1-R_0\rvert\sqrt{N}\gg 1.\label{crit}
\end{equation}

\section{The threshold case $R_0=1$}\label{sec:thresh}

We have shown above that in the limit $R_0\to 1$, the WKB solutions break down. In this section we study the behavior of the system at $R_0=1$ for both the IBM and the phenomenological model, by going beyond the WKB approach. We start again with the IBM (\ref{gsys}).  Equation (\ref{rel}) implies that $R_0=1$ corresponds to $\mu=\gamma$.
As a result, the mean-field equation, Eq.~(\ref{mf2}),  reads
\begin{equation}
\frac{d\rho}{dt}=\frac{\lambda }{2}\rho^{2}-\frac{\beta}{6}\rho^{3}.
\label{mf4}
\end{equation}
This type of equation is also encountered in combustion processes \cite{MeLl96}. Even though the linear birth and death processes cancel out at the deterministic level and do not affect the mean-field dynamics, they significantly impact the stochastic dynamics. The exact MTE of the underlying microscopic IBM,  Eq.\ (\ref{gsys}), is given by Eq.~(\ref{taum}) with the transition rates
\begin{subequations}
	\label{trans4}
	\begin{align}
	w_+(n)&=\frac{\lambda}{2}n(n-1)+\mu n,\\
	w_{-}(n)&=\frac{\beta}{6}n(n-1)(n-2)+\mu n,
\end{align}
\end{subequations}
obtained from Eqs.~(\ref{trans}) with $a=b=c=1$.

\begin{figure}[htbp]
	\includegraphics[width=\hsize]{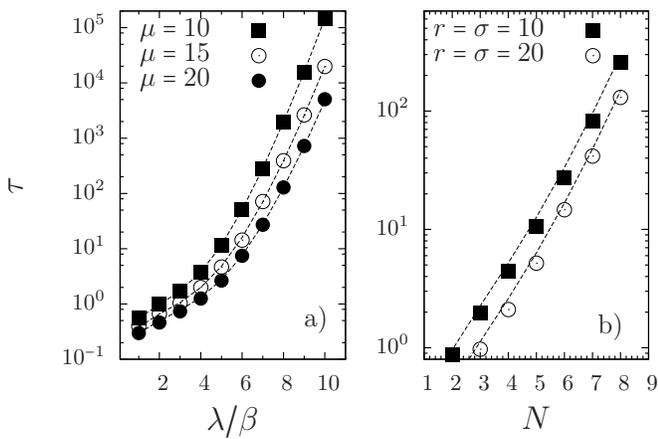}
	\caption{The MTE for $R_0=1$ is plotted as function of the characteristic population size. In panel (a) we plot the results for the
IBM with
$\mu=\gamma$ and $\beta=1$. Symbols correspond to numerical simulations of the stochastic process given by Eqs.~(\ref{gsys}). Solid lines are the results of the recursive solution for $\tau$ [Eq.~(\ref{taum})] with rates (\ref{trans4}). In panel (b) we plot the results for the phenomenological model for $r=\sigma$ and $\theta =1$. Symbols correspond to numerical simulations of the processes (\ref{gsysmodel2}) with rates (\ref{model2}). Solid lines are the results of the recursive solution (\ref{taum}) with rates (\ref{model2rates}).}
	\label{fig:f7}
\end{figure}

In Fig.~\ref{fig:f7} we compare the exact recursive solution [Eq.~(\ref{taum})] with the MTE obtained from numerical simulations
for both the IBM and phenomenological model. We have employed the Gillespie direct method \cite{Gi76,Gi77,GiHePe13}, which in this case can be shown to be more efficient than other algorithms such as the first reaction or tau-leaping algorithms  \cite{Gi01,CaGiPe06,CaLiPe04}.
Figure \ref{fig:f7} shows that for the IBM, although the parameter $\mu$ does not appear at the deterministic level [Eq.~(\ref{mf4})], it strongly influences the stochastic dynamics, and in particular, the MTE. Indeed,
the MTE decreases as $\mu=\gamma$ increases. This is because as $\mu=\gamma$ increases, the typical fluctuations grow as well. In other words, the overall noise increase causes the MTE to decrease. Note that the fact that the typical noise increases does not produce any signature at the deterministic level. The same behavior occurs in the phenomenological model as $r=\sigma$ is increased, see Fig.~\ref{fig:f7}.

Finally, it is interesting to explore the special case of $\mu=\gamma=0$. The mean-field dynamics, Eq.~(\ref{mf4}), remains unchanged, but the underlying microscopic IBM now reads,
\begin{subequations}
	\label{gsys5}
	\begin{align}
	2\,\text{X} & \xrightarrow{\lambda}  3\,\text{X},\\
	3\,\text{X} & \xrightarrow{\beta}  2\,\text{X}.
	\end{align}
\end{subequations}
In this case, the population does not go extinct since it never reaches the absorbing state at $n=0$. Instead, the system reaches a stationary state as $t\to\infty$. To determine the population abundance distribution $P(n)$ in the stationary state we set the time derivative in Eq.\ (\ref{eq:me30}) to zero.
We substitute $\mu=\gamma=0$ into the right hand side of Eq.~(\ref{eq:me30}) and obtain the recursive equation
\begin{equation}
P(n)=\frac{3\lambda}{\beta}\frac{P(n-1)}{n},\label{ab}
\end{equation}
where we set $P(0)=P(1)=0$, since the population never reaches those states. The recursive equation (\ref{ab}) can be solved in a straightforward manner for $n\geq 2$, and we obtain
\begin{equation}
P(n)=2P(2)\frac{(3\lambda/\beta)^{n-2}}{n!},
\end{equation}
where $P_2$ is a constant that can be found from the normalization condition $\sum_{n=0}^{\infty}P(n)=1$. The final result for the population abundance distribution reads
\begin{equation}
P(n)=\frac{(3\lambda/\beta)^{n}/n!}{e^{3\lambda/\beta}-1-3\lambda/\beta}.
\label{abun2}
\end{equation}

\begin{figure}[htbp]
	\includegraphics[width=\hsize]{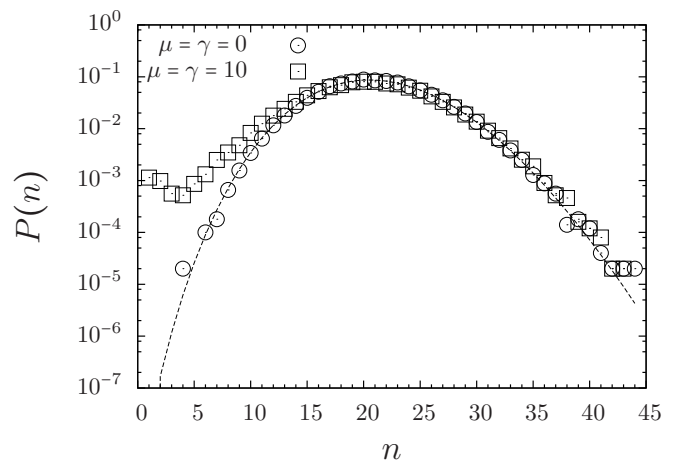}
	\caption{The population abundance distribution is plotted
	as a function of $n$ for $\lambda/\beta =7$ and $\beta=1$. The solid line is the analytical solution (\ref{abun2}), the circles are
numerical simulation results of the reaction set (\ref{gsys5}),
while the squares are numerical simulation results of the reaction set (\ref{gsys}). While the distributions' mean approximately coincides, the probability to have a small population size $n=O(1)$ becomes larger as $\mu=\gamma$ is increased, see text.}
	\label{fig:f8}
\end{figure}

We have simulated the stochastic process (\ref{gsys5}).
Once the system has relaxed to its stationary state, we have
determined the frequency of different population sizes for many realizations, obtaining $P(n)$ numerically.
In Fig.\ \ref{fig:f8} we compare the analytical result of the population abundance distribution, Eq.~(\ref{abun2}), with the result from
the numerical simulations according to the reactions~(\ref{gsys5}).
In this figure we also plot the population abundance distribution for
$\mu=\gamma= 10$, according to the reactions given by Eq.~(\ref{gsys}).
In this case, the stationary state of the system is extinction, and we sample the frequency of different population sizes at times much larger than the system's relaxation time, but much smaller than the MTE. While the mean-field equations of (\ref{gsys}) and (\ref{gsys5}) coincide, which indicates that the mean of the population abundance distributions coincides regardless of the value of $\mu=\gamma$, see Fig.\ \ref{fig:f8}, the probability to have $n=O(1)$ number of individuals is markedly different. Indeed, as $\mu=\gamma$ is increased, the left tail becomes higher owing to the fact that the typical noise strength is larger, as also demonstrated in Fig.~\ref{fig:f7}.

\section{Conclusions}\label{sec:concl}

We have investigated the stochastic dynamics of populations
with Allee effect. To that end, we have proposed two stochastic models, an individual based model and a phenomenological model with
density-dependent transition rates.
Both models are based on Markovian birth-and-death processes and are able to display the weak as well as the strong Allee effect. At the deterministic level these models are well known in the literature. However, the stochastic version of these models has not been studied. For both models we have defined a parameter $R_0$ that controls whether the Allee effect is strong, $R_0<1$, or weak, $R_0>1$.

We have employed the WKB approach to find approximate expressions for the MTE. Such approximation methods allow one to obtain accurate results for general multi-step birth-death processes, where no general exact analytical results are known. Our focus on analyzing the single-step versions of both types of models, namely the IBM  and the phenomenological model, was prompted by the fact that general exact solutions are known for this class of processes. In this way, we have been able to compare the approximation results with exact solutions, in order to assess the quality and range of validity of the WKB approach.  We have determined the dependence of the MTE on parameters of the system, in particular the dependence on $R_0$. We have also determined the relative error of the WKB results compared with the exact results for the MTE.

It was shown previously that the WKB approach breaks down in the limit of $R_0\to 1$, where the birth and death rates are equal for small populations \cite{AsMe10,EsKa09}. Here we have found an explicit criterion that
$R_0$ must satisfy, namely $\lvert R_0-1\rvert\gg 1/\sqrt{N}$, in order for the overall perturbative approach to be valid.
Furthermore, we have explored the stochastic population dynamics for this particular limit, $R_0\to 1$, by comparing the exact solution with results from numerical simulations of the underlying stochastic process. We have shown that while the mean-field dynamics is independent of the value of the birth and death rates for small populations in this limit, this is not the case for the stochastic dynamics. In fact, we have demonstrated that by influencing the typical noise strength of the system, the value of the birth and death rates for small populations strongly influences the MTE as well as the population abundance distribution prior to extinction.

\section*{Acknowledgments}
This research was partially supported by Grant No. CGL2016-78156-
C2-2-R (V.M., D.C.).

\bibliography{stochallee}
\end{document}